\documentclass[prb,preprint,showpacs]{revtex4}
\usepackage{graphics}
\begin{document}
\title{Giant enhancement of room-temperature magnetoresistance in
La$_{0.67}$Sr$_{0.33}$MnO$_{3}$/Nd$_{0.67}$Sr$_{0.33}$MnO$_{3}$ multilayers}
\author{Soumik Mukhopadhyay}
\author{I. Das}
\affiliation{Saha Institute of Nuclear Physics, 1/AF, Bidhannagar,
Kolkata 700064, India}
\begin{abstract}
The metal-insulator transition temperature in CMR manganites has been 
altered and brought close to the room temperature by 
preparing La$_{0.67}$Sr$_{0.33}$MnO$_{3}$ (LSMO)/
Nd$_{0.67}$Sr$_{0.33}$MnO$_{3}$ (NSMO)
multilayers with ultra thin individual layers of LSMO and NSMO. 
The LSMO/NSMO multilayers with ultra thin individual layers of thickness of
about $10\AA$ exhibits $150\%$ magnetoresistance at $270$ K whereas 
LSMO/NSMO multilayers with moderate 
individual layer thickness of about $40\AA$ each exhibits a mere $15\%$
magnetoresistance at the same temperature.
We have shown that the reduction in
thickness of the individual layers leads to increased spin fluctuation
which results in the enhancement of magnetoresistance.   
\end{abstract}
\pacs{73.21.Ac, 73.43.Qt, 75.47.-m, 75.47.Lx}
\maketitle

\noindent
Ever since the discovery of colossal magnetoresistance (CMR) in perovskite 
manganites~\cite{Helmholt}, extensive research on the magneotransport 
properties of manganite films, multi-layers, tunnel junctions etc. has
been initiated due to its potential for technological applications. 
The CMR manganites like La$_{0.67}$Sr$_{0.33}$MnO$_{3}$ (LSMO), 
La$_{0.67}$Ca$_{0.33}$MnO$_{3}$ (LCMO), Nd$_{0.67}$Sr$_{0.33}$MnO$_{3}$ (NSMO) 
exhibit transition from a high temperature paramagnetic insulator to a low 
temperature ferromagnetic metal. In the ground state, these manganites are 
fully spin polarized~\cite{Viret} at the Fermi level.
Encouraged by this high spin polarization, numerous efforts have been made
to achieve room temperature MR using the extrinsic magnetoresistive properties
such as the tunneling magnetoresistance (TMR)~\cite{Viret} 
phenomenon observed in manganite tunnel 
junctions or polycrystalline manganites. However it has been observed
that for manganite tunnel junctions, the TMR falls off rapidly with 
increasing temperature~\cite{Sun} 
and generally vanishes even before reaching the room temperature.
Recently large room temperature magnetoresistance has been achieved
in magnetic tunnel junctions with MgO as the tunnel barrier and Fe or CoFe
as the electrodes~\cite{yuasa,parkin}. However, in such cases the 
insulating layer MgO needs to be a highly oriented single crystal, 
which is very difficult from the fabrication point of view. Moreover,
the possibility of the transition metal electrodes having an oxidized 
and amorphous under-layer cannot be ruled out. Coming back to the 
half-metallic CMR manganites, although such materials produce high 
intrinsic MR around the Curie temperature,
the fact that their critical temperatures are well away from the room 
temperature (for LSMO $T_{c} \sim 360$ K ; for NSMO $T_{c} \sim 220$ K; for 
LCMO $T_{c} \sim 250$ K) comes in the way of obtaining large MR around 
room temperature. 

\indent 
We have studied the magneto-transport properties of LSMO/NSMO multi-layers 
where the thickness of alternate LSMO and NSMO layers have been varied. 
This study attempts to bring the metal-insulator transition temperature, 
which is usually nearly coincident with the curie temperature, closer 
to the room temperature so that large MR can be obtained around the room 
temperature. We have obtained huge enhancement in magnetoresistance near room 
temperature by reducing the thickness of alternate layers in LSMO/NSMO 
magnetic multi-layer.

\indent
Four sets of samples have been prepared. Sample $1$ is a LSMO film of thickness 
$3000\AA$ nm; sample $2$ is a NSMO film of same thickness $3000\AA$ nm; sample $3$ 
is the LSMO/NSMO multilayer with alternate layer thickness $10\AA$ 
[LSMO($10\AA$)/NSMO($10\AA$)] 
and sample $4$ is another LSMO/NSMO multilayer with alternate layer thickness 
of $40\AA$ [LSMO($40\AA$)/NSMO($40\AA$)] and with LSMO as the top layer.
Sample $4$ will 
be used from now on as a reference. The total thickness of all the 
samples is about $3000\AA$ nm each. All the samples were prepared by pulsed laser 
deposition under identical conditions and deposited on LaAlO$_{3}$ substrate.
X-ray diffraction confirms the epitaxial nature of the samples.
The magnetotransport properties were studied using 
standard four probe method. The magnetic field was applied parallel to the 
applied electric field.

\indent 
For sample $3$, the resistivity curve in absence of magnetic field 
(Fig:~\ref{fig:res}) shows a distinct peak associated with metal-insulator 
transition at around $280$ K. Considering the $T_{C}$ of bulk LSMO and 
NSMO, this is a remarkable shift in transition temperature. 
Sample $1$ and sample $4$
show more or less the same feature with no metal-insulator transition 
below $300$ K whereas sample $2$ exhibits metal insulator
transition at $220$ K. The absolute value of resisitivities at $3$ K 
in samples $1$, $2$, $3$ and $4$ are $46\, \mu\Omega$-cm, $200\, \mu\Omega$-cm,
$3950\, \mu\Omega$-cm and $300\, \mu\Omega$-cm, respectively.
Since sample $1$ and sample $4$ exhibit almost similar
magnetotransport properties and since sample $3$ and sample $4$ are
similar samples with variation in individual layer thickness only, 
we shall primarily compare the properties of sample $3$ with that 
of sample $4$. 

\indent
The resistivity data was analyzed using a polynomial 
expansion in temperature T. We have fitted the resistivity data for sample 
$3$ and $4$ with the following function 
\[\frac{\rho(T)}{\rho(0)} = 1 + \beta T^{2} + \gamma T^{4.5}\]
In the present case, the $T^2$ term comes from thermal spin 
fluctuation~\cite{Moriya} 
(of course electron-electron scattering can also give rise to such a term 
according to Fermi liquid theory~\cite{fermi} for systems with high density
of states at the fermi level). The $T^{4.5}$ 
temperature dependence has been predicted for
electron-magnon scattering in the double exchange theory~\cite{magnon}.
But the fitting is not very satisfactory at very low temperature
where the resistivity is almost temperature independent with a very minor
rise with lowering of temperature below $5$ K. We also
tried to fit with $T^{2.5}$ term instead of $T^{4.5}$ term to include 
electron-phonon scattering instead of electron magnon scattering but 
no improvement was observed. 
Henceforth we excluded the $T^{4.5}$ and $T^{2.5}$ term 
and the resistivity data was fitted considering the function
\[\frac{\rho(T)}{\rho(0)} = 1 + \beta T^{2}\] 
in the temperature range above $5$ K and below approximately $T_{MI}/2$.
Most of the studies concerning transport properties of thin films of
manganites have observed a dominant $T^{2}$ term at low temperature~\cite{electron},
which is attributed to electron-electron interaction.
Chen et. al.~\cite{chen} have fitted the ferromagnetic metallic region of the
resistivity curve using small polaronic transport term $\sinh (C/T)$, C being 
a constant and a spin wave
scattering term $T^{3.5}$. But in our case, the fitting is evidently
poor when those terms are incorporated.
To ascertain whether $T^{2}$ term is due to the predominance of spin 
fluctuations, we fitted the resistivity 
data for sample $3$ and $4$ with the above polynomial at different magnetic 
fields and studied the 
variation in $\beta$. The coefficient $\beta$ decreases substantially with 
increasing magnetic field (Fig:~\ref{fig:coeff}). This suggests the 
suppression of spin fluctuation by applied magnetic field. The observed 
decrease in $\beta$ with increasing magnetic field is more pronounced for
sample $3$ in comparison with sample $4$. This indicates that the reduction in 
thickness of individual layers has resulted in enhanced spin fluctuation
in sample $3$ compared to sample $4$.       

\indent
The magnetic field dependence of magnetoresistance (MR) 
[$\{\rho(H)-\rho(0)\}/\rho(H) (\%)$] for sample $3$ and $4$ at $3$ K shows
contrasting behavior in the low and high magnetic field region 
(Fig:~\ref{fig:mrh}: Inset). In the low magnetic field region, a sharp drop in 
resistivity with increasing magnetic field, associated with the suppression of 
domain wall scattering, is observed for sample $4$ whereas sample $3$ shows 
no low field magnetoresistance. In the high field region the MR for sample $4$
almost saturates, whereas the MR for sample $3$ exhibits linear magnetic field 
dependence. However, near the metal insulator transition temperature, 
enhancement of low field MR has been observed for sample $3$ as compared 
to sample $4$. About $8\%$ MR around $270$ K has been observed 
for sample $3$ in a magnetic field of $5$ kOe
compared to a mere $1\%$ for sample $4$. The comparison of the temperature 
dependence of high field MR between sample $1$, $3$ and $4$ is shown in
Fig:~\ref{fig:mrt}. 
Enhancement of high field MR for sample $3$ as compared to sample 
$1$ and $4$ is observed over the entire 
temperature range. For sample $3$, the MR peaks at $270$ K and then decreases 
with increasing temperature whereas the MR for sample $1$ and $4$ weakly 
increases up to $300$ K, indicating that both the samples are still well 
below the ferromagnetic transition temperature. In contrast to about $15\%$ MR at 
$270$ K for samples $1$ and $4$, the MR for sample $3$ 
at $270$ K is about $150\%$ in $80$ kOe magnetic field. The MR peak at $270$ K 
suggests that the curie temperature is very close to the metal-insulator 
transition temperature for sample $3$ and that the magnetoresistive 
properties exhibited by sample $3$ is purely intrinsic in nature. 
The temperature dependence of magnetization for sample $3$ shows that
indeed the curie temperature is at around $270$ K (Fig:~\ref{fig:mrt},inset). 
The separate transition 
temperatures at about $210$ K and $350$ K 
for the NSMO and LSMO layers are evident from the 
temperature dependence of magnetization for sample $4$ 
(Fig:~\ref{fig:mrt},inset).
At $300$ K, the MR for sample $3$ is about
$75\%$ whereas it is only about $25\%$ for samples $1$ and $4$.
We have also studied the magnetic field dependence of MR for 
sample $3$ and $4$ (Fig:~\ref{fig:mrh}) at $300$ K.
While the MR of sample $4$ shows linear magnetic field dependence, the 
MR for sample $3$ shows distinct $-H^{2}$ dependence up to about 30kOe.
This suggests that the origin of enhanced MR in sample $3$
is the increased spin fluctuation due to reduction in 
thickness of individual layers and the eventual suppression of spin 
fluctuation by external magnetic field as already enunciated at the beginning 
of the discussion.

\indent
To summarize, we have compared the magnetotransport properties between the 
two LSMO/NSMO multi-layers : LSMO($10\AA$)/NSMO($10\AA$) and 
LSMO($40\AA$)/NSMO($40\AA$), fabricated by us. The magnetotransport 
properties observed in both the multi-layers do not mimic the extrinsic 
inter-granular transport 
properties shown by ferromagnetic metal-insulator composites but are 
rather intrinsic in nature. Analyzing the resistivity data in the presence as 
well as in absence of magnetic field, we conclude that the 
reduction in thickness of the individual layers leads to increased spin 
fluctuation in LSMO($10\AA$)/NSMO($10\AA$) and the enhanced 
magnetoresistance is a consequence of suppression of spin fluctuation 
by applied magnetic field. Although no extrinsic magnetoresistive properties 
due to scattering by domain walls or spin polarized tunneling across grain 
boundaries has been observed, still enhancement of low field MR has also been 
achieved over a wide temperature range near room temperature for sample $3$.
At $270$ K and $80$ kOe magnetic field, LSMO($10\AA$)/NSMO($10\AA$)
shows $150\%$ magnetoresistance which becomes $75\%$ at $300$ K.
The enhancement of high field magnetoresistance has been observed in 
LSMO($10\AA$)/NSMO($10\AA$) over the entire temperature range compared 
to LSMO($40\AA$)/NSMO($40\AA$) multi-layer and LSMO film. 
The results suggest that it is 
possible to achieve much higher MR around room temperature by tuning the 
individual layer thickness and exploring suitable materials.

\newpage
\begin{figure}
\caption{Resistivity normalized at $3$ K, as a function of temperature, are shown
for all the samples. (The absolute values of resistivity at $3$ K are
mentioned in the text.)
The dotted curve corresponds to sample $2$, showing the $T_{MI}$ at $220$ K;
sample $1$ and sample $4$ show similar temperature dependence of
resistivity; whereas $T_{MI}$ for sample $3$ is at $280$ K, nearabout room
temperature.}
\label{fig:res}
\caption{The resistivity curves in absence
of magnetic field for samples $3$ and $4$ have been 
fitted using spin fluctuation model, i.e.
$\frac{\rho(T)}{\rho(0)} = 1 + \beta T^{2}$
Variation of $\beta$ with applied magnetic field for sample $3$ and
$4$ is shown; relative decrease of $\beta$ with applied magnetic field is greater for
sample $3$ compared to sample $4$. Inset: The $T^{2}$ dependence of resistivity
for sample $3$ is shown by the continuous line.}
\label{fig:coeff}
\caption{Temperature dependence of magnetoresistance at $80$ kOe for samples
$1$, $3$ and $4$. Sample $1$ and $4$ exhibit almost identical temperature
dependence. The MR for sample $3$ at $270$ K is about ten times compared
to that of sample $1$ and $4$. Inset: Field cooled M vs. T curves at
H = $1$ kOe for sample $3$ and sample $4$.}
\label{fig:mrt}
\caption{Magnetic field dependence of MR for sample $3$ and sample $4$ at
$300$ K. MR shows distinct $-H^{2}$ dependence up to $30$ kOe for sample $3$.
Dotted line is the theoretical fit. Inset: Magnetoresistance as a
function of magnetic field at $3$ K for sample
$3$ and sample $4$. The small low field magnetoresistance for sample $4$ is due
to suppression of domain wall scattering. In contrast, sample $3$ shows no
low field magnetoresistance and almost linear magnetic field dependence.}
\label{fig:mrh}
\end{figure}
$~~~~~~~~~~~~$
\newpage
$~~~~~~~~~~~~~~~~~~~~~~~~$
\begin{figure}
\resizebox{7.5cm}{6cm}
{\includegraphics {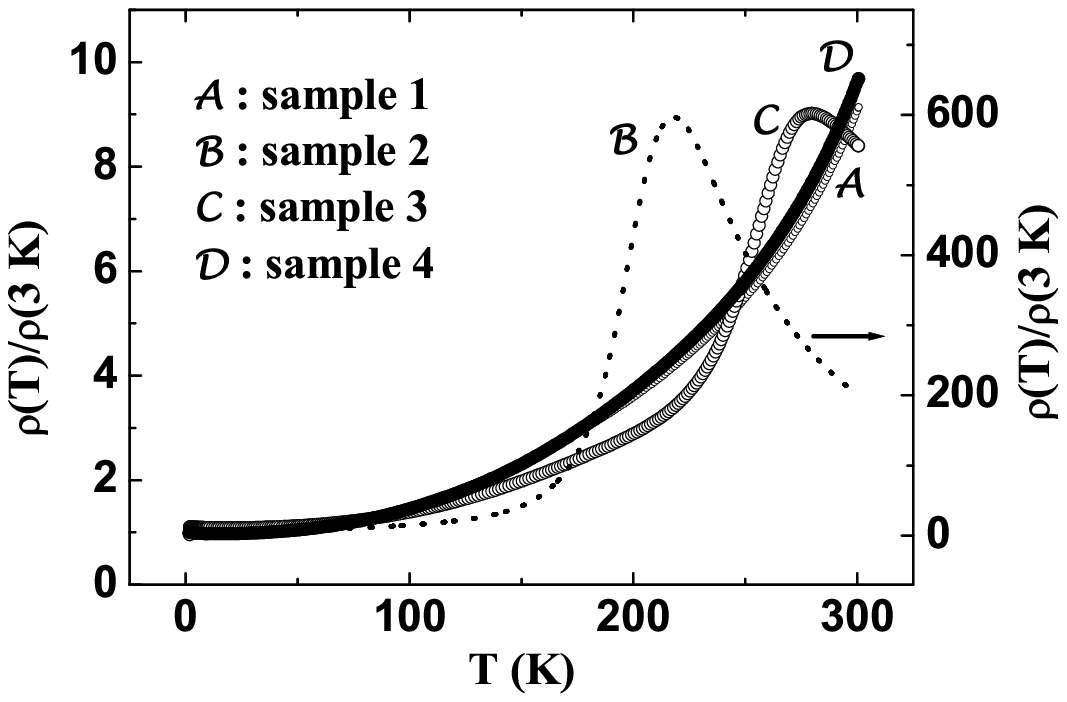}}
\end{figure}
\centering{Fig. 1}
\newpage
$~~~~~~~~~~~~~~~$
\begin{figure}
\resizebox{7.5cm}{6cm}
{\includegraphics {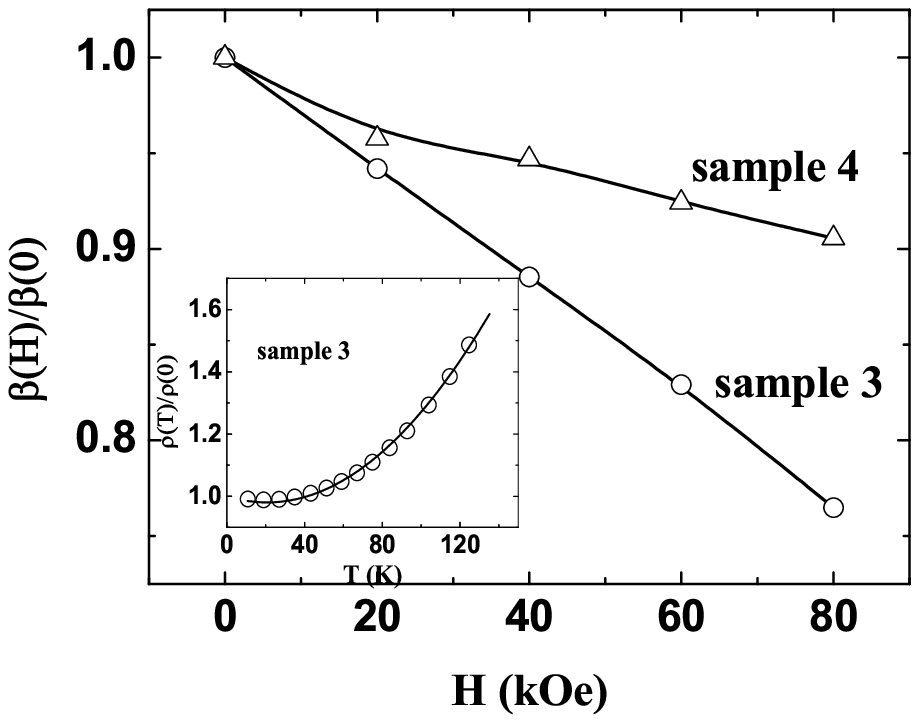}}
\end{figure}
\centering{Fig. 2}
\newpage
$~~~~~~~~~~~~~~~$
\begin{figure}
\resizebox{8.5cm}{7.5cm}
{\includegraphics {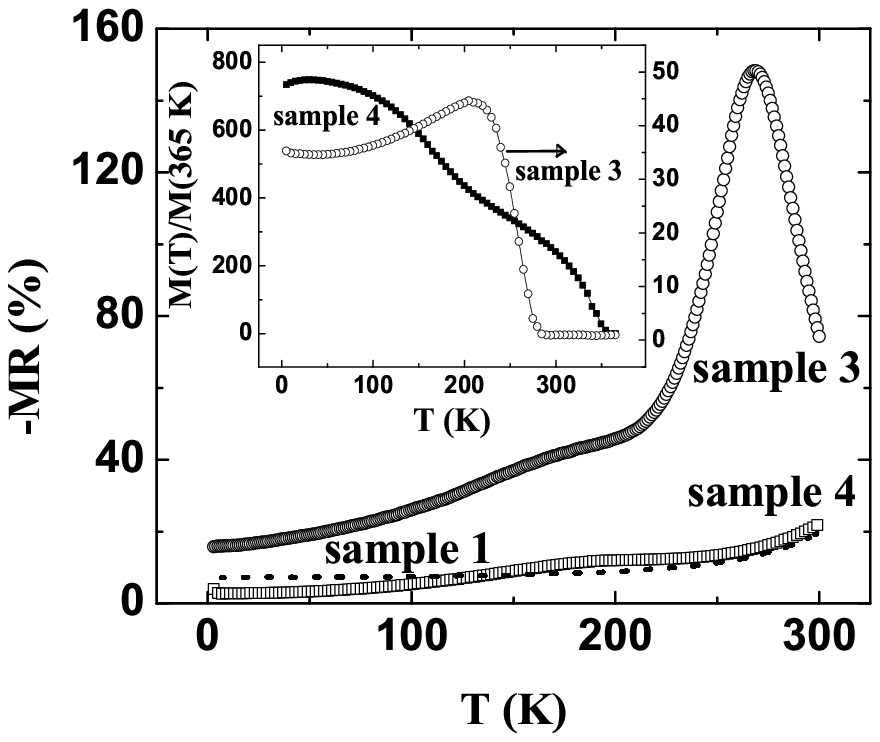}}
\end{figure}
\centering{Fig. 3}
\newpage
$~~~~~~~~~~~~~~~$
\begin{figure}
\resizebox{7.5cm}{6cm}
{\includegraphics {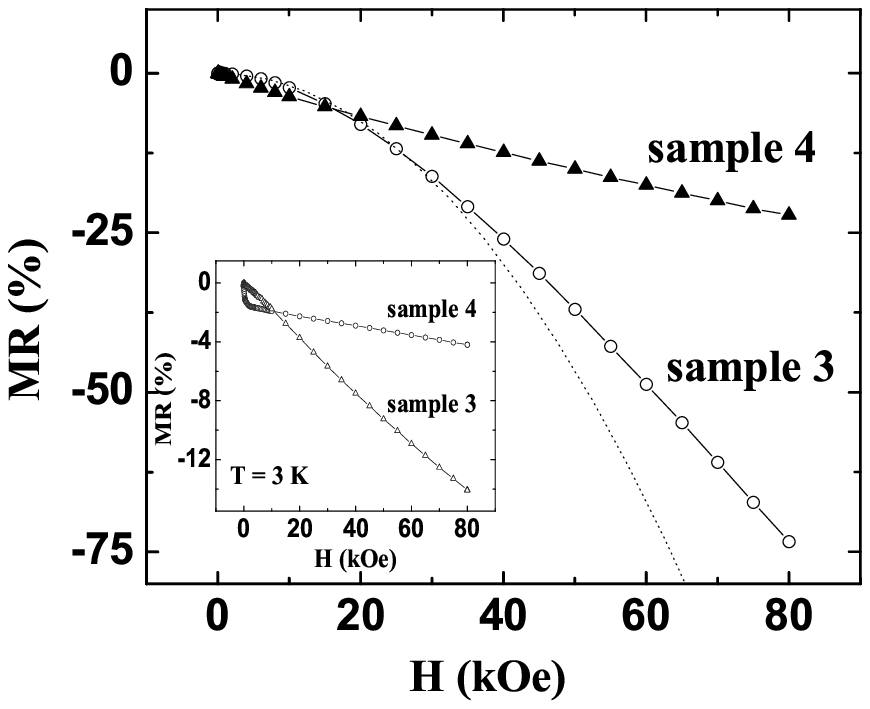}}
\end{figure}
\centering{Fig. 4}
\end{document}